\documentclass[11pt]{article}
\usepackage[margin=0.75in]{geometry}
\usepackage[all,cmtip]{xy}
\usepackage{pifont}
\usepackage{xcolor}
\definecolor{green}{rgb}{0,0.45,0}
\usepackage{amssymb}
\usepackage{geometry}
\usepackage{amsmath}
\usepackage{graphicx}
\usepackage{epstopdf}
\usepackage{amsthm}
\usepackage{lineno}
\usepackage[final]{changes}

\usepackage{authblk}
\usepackage{dsfont}
\usepackage[plain]{algorithm2e}
\newcommand{\rd}{\mathrm{d}}
\newcommand{\om}{\text{O}}
\newcommand{\jm}{\text{J}}
\newcommand{\tm}{\text{T}}
\newcommand{\tj}{\text{TJ}}
\newcommand{\jt}{\text{JT}}
\newcommand\longdashv{\mathrel{\relbar\mkern-4mu\relbar\mkern-9mu
    \vcenter{\hbox{$\dashv$}}}}
\vspace{-2cm}
\date{August 16, 2023}
\begin{document}
\title{\bf Order-of-mutation effects on cancer progression: models for
  myeloproliferative neoplasm}
\author[1,2]{Yue Wang}

\author[3,4]{Blerta Shtylla}

\author[1,5]{Tom Chou}

\affil[1]{Dept. of Computational Medicine, UCLA, Los
  Angeles, CA 90095}
\affil[2]{Irving Institute for Cancer Dynamics and Department of
  Statistics, Columbia University, New York, NY 10027}
\affil[3]{Mathematics Department, Pomona College, Claremont, CA, 91711}
\affil[4]{Quantitative Systems Pharmacology, Oncology, Pfizer, San Diego, CA 92121}
\affil[5]{Dept. of Mathematics, UCLA, Los Angeles, CA 90095\\
}
\affil[ ]{}
\affil[ ]{\textit{yw4241@columbia.edu, shtyllab@gmail.com, tomchou@ucla.edu}}
      
\maketitle{}


\begin{abstract}
%
In some patients with myeloproliferative neoplasms (MPN), two genetic
mutations are often found, JAK2 V617F and one in the TET2
gene. Whether or not one mutation is present will influence how the
other subsequent mutation affects the regulation of gene expression.
When both mutations are present, the order of their occurrence has
been shown to influence disease progression and prognosis.  We propose
a nonlinear ordinary differential equation (ODE), Moran process, and
Markov chain models to explain the non-additive and non-commutative
mutation effects on recent clinical observations of gene expression
patterns, proportions of cells with different mutations, and ages at
diagnosis of MPN. These observations consistently shape our modeling
framework. Our key proposal is that bistability in gene expression
provides a natural explanation for many observed order-of-mutation
effects.  We also propose potential experimental measurements that can
be used to confirm or refute predictions of our models.
\begin{flushleft}
\end{flushleft}
\end{abstract}

\section{Introduction}
Different genetic mutations are found in patients with
myeloproliferative neoplasm (MPN), a cancer of the bone marrow. These
mutations are known to have different effects on cell behavior
\cite{levine2007jak,delhommeau2009mutation,klampfl2013somatic,nangalia2013somatic}. In
this paper, we focus on two common mutations in MPN, JAK2 V617F
(henceforth abbreviated as JAK2) and TET2.

JAK2, the Janus kinase 2, mediates cytokine signaling to control blood
cell proliferation, while the TET2 protein catalyzes oxidation of
5-methylcytosine, thereby epigenetically influencing expression of
other genes.  It has been shown \textit{in vitro}, that JAK2 and TET2
mutations each confer a competitive growth advantage in some myeloid
cells \cite{baik2021genome,chiba2017dysregulation}. Once the JAK2 or
TET2 mutation appears in certain myeloid cells, such cells will have
higher proliferation rates than myeloid cells without such
mutations. This growth advantage is ``modest'' and might take years to
manifest itself as an increased proportion of cells carrying these
mutations in the total cell population. Thus, it is common to find
cells in a patient with different numbers of mutation types.
Moreover, in certain patients with \textit{both} JAK2 and TET2
mutations, it is possible to infer which mutation appears first.

Ortmann et al. \cite{ortmann2015effect} reported that different
mutational patterns (including the order of mutations) in
hematopoietic cells and progenitor cells are related to differences in
gene expression patterns, clonal evolution, and even macroscopic
properties. Specifically, a mutation can differentially regulate gene
expression by different amounts depending on whether or not another
type of mutation preceded it. Therefore, the change in gene expression
level when one mutation appears cannot simply be added. We call such
phenomena ``\textbf{non-additivity}''.  Additionally, patients in
which the JAK2 mutation appears before the TET2 mutation, have
different gene expression levels, percentage of cells with only one
mutation, and age at diagnosis than patients in which the TET2
mutation appears before the JAK2 mutation. This observation implies
that the order of the first appearance of these two mutations
matters. We describe such phenomena as ``\textbf{non-commutative}''.

In Section~\ref{rev}, we summarize the clinical observations reported
by Ortmann et al. \cite{ortmann2015effect}. In Section~\ref{pmom}, we
summarize previous models for such clinical observations
\cite{ortmann2015effect,swanton2015cancer,roquet2016synthetic,kent2017order,ascolani2019modeling,clarke2019using,talarmain2022hoxa9,talarmain2021modelling,mazaya2020effects,teimouri2021temporal}
and compare these models with our new models. We then build nonlinear
ordinary differential equation (ODE) models to explain the
observations regarding gene expression and list experimental evidence
that supports our models in Section~\ref{na}. In Section~\ref{at}, we
present a generalized Moran process model and three different
mechanisms to explain the observations regarding clonal evolution and
ages at diagnosis. We conclude with some discussion in
Section~\ref{disc}. A more detailed review of previous models is given
in Appendix~\ref{review}, while an alternative, but related Markov
chain model to explain non-commutative effects of mutations on gene
expression is presented in Appendix~\ref{mcm}.

\section{Clinical observations on the effects mutation order}
\label{rev}

For patients exhibiting cells with \textit{both} JAK2 and TET2
mutations, one might ask: Which mutation occurred first in the
patient? If we find cells with only JAK2 mutations, cells with both
JAK2 and TET2 mutations, but no cells with only the TET2 mutation,
then the JAK2 mutation must have appeared in the patient before the
TET2 mutation.  Such patients are classified as JAK2-first.  Patients
in which we find doubly mutated cells and TET2-only cells but not
JAK2-only cells are classified as TET2-first.  If a patient carries
JAK2-only cells, TET2-only cells, and JAK2-TET2 cells, then both JAK2
and TET2 mutations occurred independently in wild-type cells and more
information, such as other associated mutations or tagging that
resolves subpopulations, is needed to infer their temporal order of
appearance. Such patients were not considered by Ortmann et
al. \cite{ortmann2015effect}. For more complex samples that contain
cells with multiple types of mutations, one can use different
algorithms to determine the probabilities of different orders of
mutations from sequencing data
\cite{de2018single,pellegrina2022discovering,ramazzotti2019learning,khakabimamaghani2019uncovering,gao2022phylogenetic}. However,
patients with ambiguous cell populations (JAK2-only cells, TET2-only
cells, and JAK2-TET2 cells) were not considered by Ortmann et
al. \cite{ortmann2015effect}.

Besides inferring the order of mutations, Ortmann et
al. \cite{ortmann2015effect} also measured bulk gene expression levels
from MPN-patient-derived populations of cells containing different
sets of mutations.  Their observations are summarized in Table
\ref{notation} in which $x^*$ denotes the steady state expression
level of gene X in a cell and the subscripts define mutation status of
the cell.


\begin{table}[]
\begin{center}
\begin{tabular}{|r|c|c|}
\hline
 \:  & w/o JAK2 mutation & with JAK2 mutation   \\
\hline
w/o TET2 mutation &   $x^*_\om$  &  $x^*_\jm$   \\
\hline
with TET2 mutation    &   $x^*_\tm$  & 
\begin{tabular}[c]{@{}l@{}} $x^*_\jt$ (JAK2-first) \\ 
$x^*_\tj$ (TET2-first)
\end{tabular} \\
\hline
\end{tabular}
\end{center}
\caption{Definition of stationary gene expression levels $x^*$ for
  cells with different mutation patterns.}
\label{notation}
\end{table}
%


\vspace{4mm}

\noindent \textbf{(1)} Some genes are up-regulated (or down-regulated)
by a JAK2 mutation only if the TET2 mutation is not present. If the
TET2 mutation is also present, gene expression is not affected. Thus
$x^*_\tm=x^*_\tj$, $x^*_\om > x^*_\jm$ or $x^*_\om < x^*_\jm$.

\vspace{3mm}
\noindent \textbf{(2)} Other genes are up-regulated (or
down-regulated) by JAK2 mutations only if TET2 mutations are also
present; but they are not affected if the TET2 mutation is
\textit{not} present. For these cases, $x^*_\om=x^*_\jm$,
$x^*_\jm>x^*_\tj$ or $x^*_\jm<x^*_\tj$.

\vspace{3mm}
\noindent \textbf{(3)} Ten genes (AURKB, FHOD1, HTRA2, IDH2, MCM2,
MCM4, MCM5, TK1, UQCRC1, WDR34) are up-regulated in cells with JAK2
mutations if TET2 mutations are not present, but they are down-regulated
by JAK2 mutations if TET2 mutations \textit{are} present.  This
scenario corresponds to $x^*_\om < x^*_\jm$, $x^*_\tm > x^*_\tj$.

\vspace{3mm}
\noindent \textbf{(4)} {Different orders of appearances of JAK2 and
TET2 mutations seem to have different effects on other genes so that
$x^*_\jt \ne x^*_\tj$.} These conclusions are inferred from
  other indirect evidence (\textit{e.g.}, JAK2-first cells are more
  sensitive to ruxolitinib than TET2-first cells \cite{ortmann2015effect}).

\vspace{4mm} 

\noindent Observations {\bf (1-3)} can be regarded as
\textbf{non-additivity}, since the effect of JAK2 mutation differs
with or without TET2 mutation.  In other words, $x^*_\jm - x^*_\om \ne
x^*_\tj - x^*_\tm$.  Observation {\bf (4)} represents
\textbf{non-commutativity} since exchanging the order of acquiring
different mutations can lead to different expression levels or cell
states \cite{levine2019roles}. Mathematically,
$x^*_\om+(x^*_\jm-x^*_\om)+(x^*_\jt-x^*_\jm)=x^*_\jt\ne
x^*_\tj=x^*_\om+(x^*_\tm-x^*_\om)+(x^*_\tj-x^*_\tm)$.  In fact, if the
gene expression levels are additive with respect to multiple mutations,
namely $x^*_\jm - x^*_\om = x^*_\tj-x^*_\tm$ and $x^*_\tm-x^*_\om =
x^*_\jt-x^*_\jm$, then it is also commutative: $x^*_\jt =
x^*_\jm+x^*_\tm-x^*_\om=x^*_\tj$.  Therefore, non-commutativity is a
special case of non-additivity. At the cell or tissue level, Ortmann
et al. \cite{ortmann2015effect} also report two observations
specifically related to non-commutativity:

\vspace{4mm}
 
\noindent \textbf{(5)} In TET2-first patients, the percentage of cells
with just one mutation (TET2) is significantly higher than the
percentage of JAK2-only cells in JAK2-first patients.

\vspace{3mm} 
\noindent \textbf{(6)} At diagnosis, JAK2-first patients are
significantly younger than TET2-first patients.

\vspace{4mm} 

Ortmann et al. \cite{ortmann2015effect} also report other observations
such as differences in MPN classification and risk of thrombosis
between JAK2-first and TET2-first patients.  These are covered by
observations \textbf{(1--6)}, particularly {\bf (4)}, and we do not
explicitly discuss them here.

\section{Comparison between previous models and our models}
\label{pmom}

There have been models put forth that explain observations {\bf
  (1-6)}. We briefly summarize previous models in this section. See
Appendix~\ref{review} for a detailed review of previous models. These
past models and our set of models are compared in Table~\ref{sum}. Our
models provide better coverage of the observed phenomena and can be
concatenated for a more complete picture of MPN progression. We now
provide an overview of our more complete analysis, filling in some
mechanistic explanations of observations {\bf (1-6)}.

\begin{table}[]
  \begin{center}
\begin{tabular}{l|l|c|l}
\hline 
{\bf Previous Studies}  & Key assumptions  & Quantitative? 
%
%
  & \begin{tabular}[c]{@{}l@{}}Observation(s)\\ 
explained\end{tabular} \\
\hline
Kent and Green \cite{kent2017order}  & \begin{tabular}[c]{@{}l@{}}JAK2 and TET2 mutations \\
compete for the same\\  regulation region\end{tabular} & {\color{red}\ding{55}}  & {\bf (4)}  \\ 
\hline
Kent and Green \cite{kent2017order}  & \begin{tabular}[c]{@{}l@{}}JAK2 and TET2 have \\ 
different effects on \\
microenvironments\end{tabular}  & {\color{red}\ding{55}}  & {\bf (4)}  \\
\hline
Roquet et al. \cite{roquet2016synthetic}   & \begin{tabular}[c]{@{}l@{}}Mutations work\\ 
like recombinases\end{tabular} & {\color{red}\ding{55}} & {\bf (4)}   \\ 
\hline
\begin{tabular}[c]{@{}l@{}}Clarke et al. \cite{clarke2019using}\\
Talarmain et al. \cite{talarmain2022hoxa9,talarmain2021modelling}\\ 
Mazaya et al. \cite{mazaya2020effects}\end{tabular}   & 
\begin{tabular}[c]{@{}l@{}}Gene expression satisfies\\ 
a generalized boolean network\\ 
model with multistability\end{tabular} & {\color{green}\ding{51}}  & {\bf (4)} \\    
\hline
\begin{tabular}[c]{@{}l@{}}Ortmann et al. \cite{ortmann2015effect}\\
  Ascolani and Li{\`o} \cite{ascolani2019modeling}\\ 
Clarke et al. \cite{clarke2019using}\end{tabular} & 
\begin{tabular}[c]{@{}l@{}}JAK2 and TET2 mutations \\ 
confer different advantages \\ 
to cell proliferation\end{tabular} & {\color{green}\ding{51}}  & {\bf (5)}  \\ 
\hline
\begin{tabular}[c]{@{}l@{}}Teimouri and \\ 
Kolomeisky \cite{teimouri2021temporal}\end{tabular}  & 
\begin{tabular}[c]{@{}l@{}}JAK2 and TET2 mutations \\ 
	bring different advantages \\ 
	to cell proliferation\end{tabular}  & {\color{green}\ding{51}}   & {\bf (6)}  \\  
\hline\hline 
    {\bf Current Analysis}  & Key assumption  & Quantitative?
%
%
 &   \begin{tabular}[c]{@{}l@{}}Observation(s)\\ 
explained\end{tabular} \\
\hline
ODE model   & \begin{tabular}[c]{@{}l@{}}Gene expression satisfies a\\ 
nonlinear ODE with bistability\end{tabular}   & {\color{green}\ding{51}}    & {\bf (1, 2, 3, 4)}  \\ 
\hline  
Markov chain model  & 
\begin{tabular}[c]{@{}l@{}}Gene expression satisfies a\\ 
Markov chain with bistability\end{tabular}  & {\color{green}\ding{51}}    & {\bf (4)} \\
\hline
\begin{tabular}[c]{@{}l@{}}Moran process, \\ 
	Mechanism {\bf (A)} \end{tabular} &
  \begin{tabular}[c]{@{}l@{}}JAK2 and TET2 mutations \\ 
bring different advantages \\ to cell proliferation\end{tabular}  
& {\color{green}\ding{51}}  & {\bf (5, 6)}   \\
\hline         
\begin{tabular}[c]{@{}l@{}}Moran process, \\ 
	Mechanism {\bf (B)}\end{tabular} & 
\begin{tabular}[c]{@{}l@{}}Mutation rates for JAK2 and \\ TET2 are different\end{tabular}  
& {\color{green}\ding{51}}  & {\bf (5, 6)}  \\  
\hline
\begin{tabular}[c]{@{}l@{}}Moran process, \\ 
	Mechanism {\bf (C)}\end{tabular} & \begin{tabular}[c]{@{}l@{}}JAK2 mutation can induce\\ 
TET2 mutation\end{tabular} & {\color{green}\ding{51}} & {\bf (5, 6)} \\
\hline
\end{tabular}
\end{center}
\caption{A summary of studies in mutational order and how
    they address observations given in \cite{ortmann2015effect}.
    Previous studies (top) and corresponding explanations are compared
    with the understanding afforded by our proposed mechanisms and
    models (bottom).}
\label{sum}
\end{table}


Observations {\bf (1)} and {\bf (2)}, the up-regulation of certain
genes depending on the presence and absence of certain mutations, form
a common ``logic gate'' in which expression levels can be
changed if and only if both conditions are met. We first construct a
simple nonlinear ODE model to explain observations {\bf (1, 2)} by
defining a threshold that is passed if and only if both conditions are
satisfied. This model will serve as a building block for an
explanation of observation {\bf (3)}, for which no model has thusfar
been proposed.  We also explain {\bf (3)} using a nonlinear ODE model
and find two candidates for a hidden factor in this model. Some
regulatory terms in our model have been verified experimentally, and
we propose experiments to examine other regulatory relationships.

Kent and Green \cite{kent2017order} explain the non-commutative
order-of-mutation effects (observation {\bf (4)}) by invoking
hypothetical mechanisms that lack experimental evidence, while Roquet
et al. \cite{roquet2016synthetic} simply proposed a mathematical space
in which operators are not commutative and not really connected to
genetic mutations.  Clarke et al. \cite{clarke2019using}, Talarmain et
al. \cite{talarmain2022hoxa9,talarmain2021modelling}, and Mazaya et
al. \cite{mazaya2020effects} all use (generalized) boolean networks to
explain observation {\bf (4)}. The state space is discrete, and the
deterministic dynamics with many parameters are chosen artificially
with little justification. At the single-cell level, gene expression
levels are discrete and stochastic, but at the bulk level it is
approximately deterministic and continuous. Therefore, we propose a
nonlinear ODE model (deterministic, continuous-state) and a Markov
chain model (stochastic, discrete-state) to describe observation {\bf
  (4)}.

Observation {\bf (5)} can be simply explained by assuming that
different mutations give rise to different proliferation advantages,
as qualitatively described in Ortmann et al. \cite{ortmann2015effect}.
Ascolani and Li{\`o} \cite{ascolani2019modeling} model driver and
passenger mutations using a similar assumption, but not for MPN.

The only possible explanation put forth for younger JAK2-first
patients at diagnosis (observation {\bf (6)}) was provided by Teimouri
and Kolomeisky \cite{teimouri2021temporal} who also assumed that the
different mutations carry different proliferation advantages. In their
model, the second mutation can appear if and only if the first
mutation rapidly expands in the cell population. Because they assume
that reaching the final state that all cells have both mutations is
conditioned on no extinction once a mutation appears, they
underestimate the predicted time to reach the final state.



To study observations {\bf (5, 6)}, we consider a generalized Moran
process, which is a more realistic model for describing the population
dynamics of hematopoietic stem and progenitor cells. We find that
three different parameter limits, describing three distinct biological
mechanisms (including that proposed by Teimouri and Kolomeisky) can
reproduce observations {\bf (5, 6)} separately.


 


\section{Models for non-additivity and non-commutativity in gene expression}
\label{na}

In this section, we build models that provide mechanistic explanations
for observations {\bf (1, 2, 3, 4)}, emphasizing the non-additive and
non-commutativity properties of two mutations on gene expression.

\subsection{Mathematical background}
First, consider ordinary differential equation (ODE) models for gene
expression and regulation. For gene X with expression level $x(t)$,
the simplest model $\rd x/\rd t=\lambda-\gamma x$ considers only synthesis and
degradation with constant rates $\lambda, \gamma$, and a stationary state
$x^*=\lambda/\gamma$. If other genes (mutations) regulate the expression of X, we
can allow the synthesis rate $\lambda$ to depend on other factors, which may
include the activation state of genes Y and Z.  For example, we might
write a deterministic model for the expression level $x(t)$ as

\begin{equation}
\frac{\rd x(t)}{\rd t}= \lambda_{0} + \lambda_\mathrm{Y}\mathds{1}_{\mathrm{Y}}
+ \lambda_\mathrm{Z}\mathds{1}_{\mathrm{Z}}-\gamma x.
\label{ODE0}
\end{equation}
Here, we have modeled the synthesis rate $\lambda = \lambda_{0} +
\lambda_\mathrm{Y}\mathds{1}_{\mathrm{Y}}+
\lambda_\mathrm{Z}\mathds{1}_{\mathrm{Z}}$ as a Boolean control operator
with $\mathds{1}_{\mathrm{Y}}=1$ if Y (gene activity or product) is
present, $\mathds{1}_{\mathrm{Y}}=0$ otherwise, and $\lambda_\mathrm{Y}$ is
a constant regulation amplitude of gene Y on the expression of gene X.
A similar term with amplitude $\lambda_\mathrm{Z}$ arises for mutation Z.
After $\mathds{1}_{\mathrm{Y}}$ or $\mathds{1}_{\mathrm{Z}}$ changes
(e.g., one gene mutates), the expression level of X will eventually return
to a new equilibrium. Therefore, in this section, we only consider
the stationary state $x^*$.

\begin{figure}[h!]
  \centering
  \includegraphics[width=0.525\linewidth]{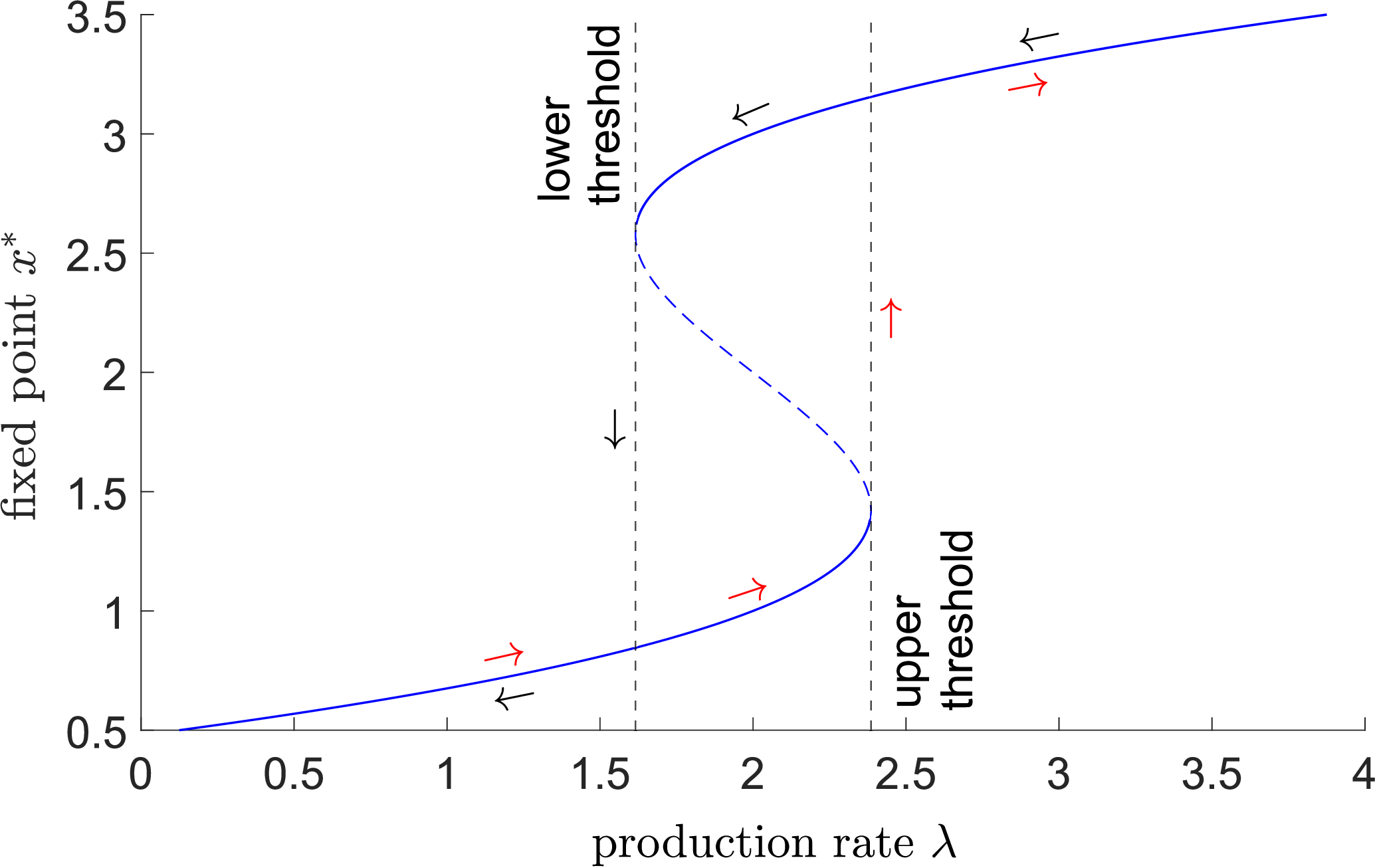}
	\caption{\small The fixed points of Eq.~\ref{eqnew} for different
          values of the expression rate of gene X, $\lambda$. The
          solid blue line is the stable fixed point, and the dashed
          blue line is the unstable fixed point. If the system starts
          at $\lambda <1.6$, the only fixed point is $x^*\approx
          0.8$. As $\lambda$ is increased, the system moves along the
          red arrow in the low branch until reaching the upper
          threshold at $\lambda\approx 2.4$, at which point the system
          jumps to the high-$x^{*}$ branch, and the fixed point jumps
          from $x^*\approx 1.4$ to $x^*\approx 3.2$. If the synthesis
          rate is lowered starting from a value $\lambda > 2.4$, the
          system moves along the black arrow along the high-value
          branch until reaching the lower threshold at $\lambda\approx
          1.6$, at which the system jumps from $x^*\approx 2.6$ to the
          low-value branch at $x^*\approx 0.8$.}
	\label{c0}
\end{figure}

The linear (in $x$) ODE in Eq.~\ref{ODE0} cannot explain observations
{\bf (1-4)} since the regulation effects of different genes
(mutations) are additive and commutative. Regardless of the status of
other genes and mutations, the presence of one mutation always has the
same effect. Therefore, one needs to include a nonlinear term.
Consider

\begin{equation}
    \frac{\rd x(t)}{\rd t}= \lambda + f(x)-x
	\label{eqnew}
\end{equation}
where for simplicity we have normalized time so that the intrinsic
degradation rate $\gamma \equiv 1$ and $\lambda$ is the
\added{dimensionless} synthesis rate that may still depend on the
presence of mutations of other genes (thus being externally tunable).
The nonlinear term $f(x)$ represents the autoregulation of $X$
\cite{wang2023inference}. A possible form of $f(x)$ is

\begin{equation}
  f(x) =-(x-2)^3+2(x-2).
  \label{eq:fx}
\end{equation}
While many possible forms for $f(x)$ may be inferred from measurements
or otherwise approximated or modeled, we will use the form in
Eq.~\ref{eq:fx} to explicitly illustrate the effects of such an
autoregulation term on the expression of X.  The fixed points
(stationary states) of Eq.~\ref{eqnew} using $f(x)$ given in
Eq.~\ref{eq:fx} are plotted in Fig.~\ref{c0} as a function of $\lambda$ and
show the high and low expression level branches.

For this nondimensionalized model, when $\lambda < 1.6$, there is one stable,
low-value fixed point at $x^{*}\lesssim 0.8$. If $\lambda >2.4$, there is one
stable fixed point $x^{*} \gtrsim 3.2$ continued from the stable high-value
branch. At intermediate values $1.6< \lambda <2.4$, both values of $x^*$
(high and low) are locally stable and are connected by an unstable
middle branch of fixed points (dashed curve).

When we start from $\lambda<1.6$, the system resides only on the low
expression level branch. If $\lambda$ is then increased to
$1.6<\lambda<2.4$, although there are two stable branches, the system
stays at the low-$x^{*}$ branch. When we further increase $\lambda$
until $\lambda>2.4$, the stable low-$x^{*}$ branch and the unstable
middle branch collide and disappear (saddle-node bifurcation), and the
system jumps to the stable high-$x^{*}$ branch. If we start with
$\lambda>2.4$, the system is at the high-level branch.  Decreasing
$\lambda$ to $1.6<\lambda<2.4$, the system will stay at the stable
high level branch until $\lambda<1.6$, when the stable high-level
branch and the unstable intermediate-value branch collide and
disappear, and the system jumps to the low-$x^{*}$ branch. In this
model, when we change the parameter $\lambda$ along different
trajectories, even though they all arrive at final values
$1.6<\lambda<2.4$, the stationary state can differ. For example, if
the value of $\lambda$ is evolved according to $\lambda=2\to
\lambda=1\to \lambda=2$, the final state is $x^*=1$, but if $\lambda$
follows the trajectory $\lambda=2\to \lambda=3\to\lambda=2$, the final
state is the high-value one at $x^*=3$.



Now consider a model in which the source of $X$ is controlled by genes
Y and Z through
$\lambda=\lambda_{0}+\lambda_\mathrm{Y}\mathds{1}_\mathrm{Y}+\lambda_\mathrm{Z}\mathds{1}_\mathrm{Z}$.
Genes Y and Z can qualitatively affect the stationary state values of
expression of X, $x^{*}$, if including their presence (or absence)
induces $\lambda$ to cross the thresholds at $1.6$ and $2.4$. This
model structure means that different orders of mutations (changes in Y
and Z) can give rise to different stationary states and lead to
non-additive and non-commutative effects on X.  We now use the model
structure given by Eq.~\ref{eqnew} to explain observations {\bf (1, 2,
  3, 4)}. Note that we just need Eq.~\ref{eqnew} to be nonlinear (to
generate non-additivity) and exhibit bistability (to induce
non-commutativity).


\subsection{Models for observations {\bf (1, 2)}}
\begin{figure}[h!]
\centering
    \includegraphics[width=0.83\linewidth]{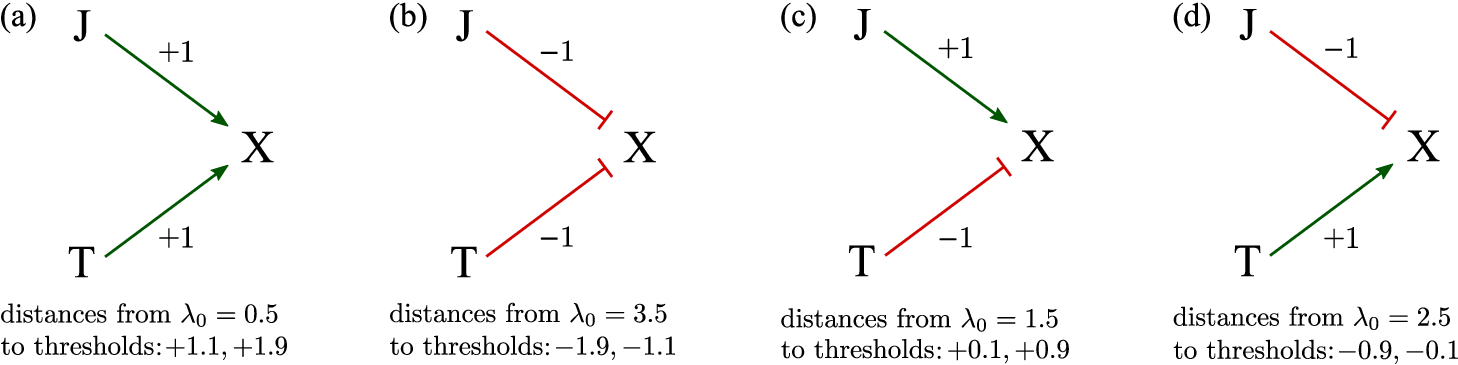}
  \caption{\small (a) A schematic of the model
    $\lambda=0.5+\mathds{1}_\mathrm{J}+\mathds{1}_\mathrm{T}$ that
    yields $x^*_\om=x^*_\jm$ and $x^*_\tm<x^*_\tj$.  ``J
    $\longrightarrow$ X'' indicates that the presence of a J mutation
    up-regulates expression of X. In this particular model, we have
    $x^*_\om\approx 0.6$, $x^*_\jm\approx 0.8$, $x^*_\tm\approx 0.8$,
    and $x^*_\tj\approx 3.2$. (b) Schematic of the model
    $\lambda=3.5-\mathds{1}_\mathrm{J}-\mathds{1}_\mathrm{T}$ which
    yields $x^*_\om=x^*_\jm$ but $x^*_\tm>x^*_\tj$. ``J,T $\longdashv$
    X'' indicates that JAK2 and TET2 mutations both down-regulate
    X. Here, we have $x^*_\om\approx 3.4$, $x^*_\jm\approx 3.2$,
    $x^*_\tm\approx 3.2$, $x^*_\tj\approx 0.8$. (c) The production
    rate model
    $\lambda=1.5+\mathds{1}_\mathrm{J}-\mathds{1}_\mathrm{T}$ captures
    $x^*_\om<x^*_\jm$, $x^*_\tm=x^*_\tj$. Here, we have
    $x^*_\om\approx 0.8$, $x^*_\jm\approx 3.2$, $x^*_\tm\approx 0.6$,
    $x^*_\tj\approx 0.8$. (d)
    $\lambda=2.5-\mathds{1}_\mathrm{J}+\mathds{1}_\mathrm{T}$ explains
    $x^*_\om>x^*_\jm$ but maintains $x^*_\tm=x^*_\tj$. As before, the
    symbols ``$\longrightarrow$'' and ``$\longdashv$'' represent
    up-regulation and down-regulation, respectively. This scenario
    yields $x^*_\om\approx 3.2$, $x^*_\jm\approx 0.8$, $x^*_\tm\approx
    3.4$, $x^*_\tj\approx 3.2$. The distances of the lower and upper
    thresholds to the value of $\lambda_{0}$ are indicated for all
    cases.}
	\label{f1}
\end{figure}

We consider different variants of Eq.~\ref{eqnew} to explain why some
genes have $x^*_\om \ne x^*_\jm$, but $x^*_\tm = x^*_\tj$ (and vice
versa). In the following, ``J'' will indicate the JAK2 mutation while
``T'' will denote the TET2 mutation. In this application, Y and Z are
identified as target genes regulated by J and T. Thus, we can simplify
the expression rate $\lambda$ in Eq.~\ref{eqnew} to, \textit{e.g.},
$\lambda=0.5+\mathds{1}_\mathrm{J}+\mathds{1}_\mathrm{T}$. With no
mutation, $\lambda =\lambda_{0}=0.5$, and the system is in the
low-expression state $x^*_\om\approx 0.6$. Consider a scenario in
which $\mathds{1}_{\mathrm{T}}=0$, $\mathds{1}_{\mathrm{J}}=1$,
\textit{i.e.}, the JAK2 mutation is present but not the TET2 mutation
(or vice versa). Then, $\lambda=1.5$ and the system is at the
low-$x^{*}$ state $x^*_\jm\approx 0.8$ (also, $x^*_\tm\approx
0.8$). If both JAK2 and TET2 mutations are present, then
$\mathds{1}_{\mathrm{T}}=\mathds{1}_{\mathrm{J}}=1$, $\lambda=2.5$,
and the system is in the high-$x^{*}$ state $x^*_\tj\approx 3.2$.  We
have $x^*_\om \approx x^*_\jm$ but $x^*_\tm < x^*_\tj$. Therefore, in
this case, JAK2 mutation up-regulates X only if the TET2 mutation is
present. See Fig.~\ref{f1}(a) for an illustration of this scenario.

Now, assume the regulated production rate takes the form
$\lambda=3.5-\mathds{1}_\mathrm{J}-\mathds{1}_\mathrm{T}$. If
$\mathds{1}_{\mathrm{T}}=0$ (no TET2 mutation), then $\lambda=3.5$ and
a JAK2 mutation itself does not affect X expression much
($x^*_\jm\approx 3.2$, $x^*_\om\approx 3.4$). If
$\mathds{1}_{\mathrm{T}}=1$, then a JAK2 mutation (changing
$\mathds{1}_{\mathrm{J}}$ from $0$ to $1$) will alter the
$x$-production rate to $\lambda=1.5$, sufficient to decrease the
steady state expression from $x^*_\tm\approx 3.2$ to $x^*_\tj\approx
0.8$. While $x^*_\om \approx x^*_\jm$, $x^*_\tm> x^*_\tj$. Thus, the
JAK2 mutation down-regulates expression of X only if the TET2 mutation
is present. See Fig.~\ref{f1}(b) for a schematic of this scenario.

Now, if $\lambda=1.5+\mathds{1}_\mathrm{J}-\mathds{1}_\mathrm{T}$, then if
the T is absent (no TET2 mutation), the presence of J (a JAK2
mutation) up-regulates X since $x^*_\om\approx 0.8$, $x^*_\jm\approx
3.2$. In the presence of T, J does not affect X expression much since
$x^*_\tm\approx 0.6$, $x^*_\tj\approx 0.8$. This regulation
model is depicted in Fig.~\ref{f1}(c).

Finally, consider a gene expression rate governed by 
$\lambda=2.5-\mathds{1}_\mathrm{J}+\mathds{1}_\mathrm{T}$, as shown in
Fig.~\ref{f1}(d).  If T is
not present, then J down-regulates X since $x^*_\om\approx 3.2$,
$x^*_\jm\approx 0.8$.  In the presence of T, J does not affect X
expression much since $x^*_\tm\approx 3.4$, $x^*_\tj\approx 3.2$.

\subsection{Model for observation (3)}
\label{p3}
\begin{figure}[h!]
\centering
    \includegraphics[width=0.65\linewidth]{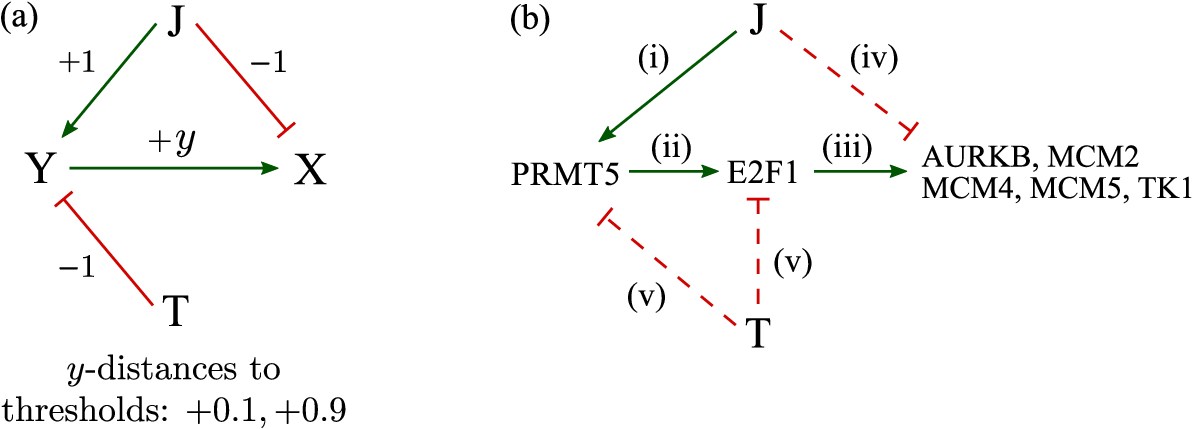}
%
        \caption{\small (a) Schematic of a model that explains
          $x^*_\om<x^*_\jm$ but $x^*_\tm>x^*_\tj$. In this model, the
          steady-state expression levels of Y are $y^*_\om\approx
          0.8$, $y^*_\jm\approx 3.2$, $y^*_\tm\approx 0.6$,
          $y^*_\tj\approx 0.8$. The basal value of $x=1$, while the
          different stationary expression levels of X are
          $x^*_\om\approx 1.8$, $x^*_\jm\approx 3.2$, $x^*_\tm\approx
          1.6$, $x^*_\tj\approx 0.8$. (b) The gene regulatory network
          that explains $x^*_\om<x^*_\jm$ but $x^*_\tm>x^*_\tj$ for a
          number of genes. Solid line indicates a verified regulation
          while the dashed line denotes a hypothesized regulatory
          interaction.}
  	\label{f5}
  \end{figure}

To explain observation {\bf (3)} that $\om \to \jm$ and $\tm \to \tj$ have
opposite effects, we need a more complicated variant of Eq.~\ref{eqnew}.
Consider a gene Y whose expression level $y$ is described by 
\begin{equation}
  \frac{\rd y}{\rd t}=\lambda+f(y)-y,
  \end{equation}
in which $\lambda=1.5+\mathds{1}_\mathrm{J}-\mathds{1}_\mathrm{T}$ and
$f(y)=-(y-2)^3+2(y-2)$. This setup gives rise to $y^*_\om\approx 0.8$,
$y^*_\tm\approx 0.6$, $y^*_\tj\approx 0.8$, and $y^*_\jm\approx 3.2$.
Now, consider a gene X whose expression level follows the linear dynamics
\begin{equation}
  \frac{\rd x}{\rd t}=1-\mathds{1}_\mathrm{J}+y-x,
  \end{equation}
where X has a basal synthesis and decay rate of $1$. A JAK2 mutation
can directly down-regulate X expression with strength $1$, while
expression of Y can up-regulate that of X with strength proportional
to its expression level $y$.
%
%
Fig.~\ref{f5}(a) shows the key regulation processes in this
model. Without JAK2 and TET2 mutations, $\lambda=1.5$, which is under
the lower threshold of $\lambda=1.6$. In this case, Y is in its
low-expression state $y^*_\om\approx 0.8$ and X is only weakly
affected by Y, with a stationary expression level $x^*_\om\approx
1.8$. With J but not T, $\lambda=2.5$, which is above the upper
threshold $2.4$. In this case, Y is in its high-value state
$y^*_\jm\approx 3.2$. Now, X expression is affected by both J and Y
(strongly), taking on the value $x^{*}_{\jm} \approx 3.2$.  With T but
not J, $\lambda=0.5$, below the lower threshold of $1.6$. In this
case, Y is in its low-value state $y^*_\tm\approx 0.6$ and X
expression, $x^*_\tm\approx 1.6$, is only weakly affected by Y
expression.

In the presence of both JAK2 and TET2 mutations, $\lambda=1.5$, under the
lower threshold of $1.6$. In this case, Y is in its low-value state
$y^*_\tj\approx 0.8$ and X is affected weakly by Y expression and by
the JAK2 mutation, with $x^*_\tj\approx 0.8$.  Therefore, without a
TET2 mutation, JAK2 mutation up-regulates X expression (from
$x^*_\om\approx 1.8$ to $x^*_\jm\approx 3.2$); with the TET2 mutation,
a JAK2 mutation down-regulates X expression from $x^*_\tm\approx 1.6$
to $x^*_\tj\approx 0.8$.

This proposed model introduces an extra gene Y in order to explain
$x^*_\om<x^*_\jm$ and $x^*_\tm>x^*_\tj$. To connect our ODE model to
observations of specific genes, and to find potential candidates for Y, we
list some specific experimental findings:

\begin{itemize}

\item[(i)] For MPN, the expression of PRMT5 is increased in cells with
  the JAK2 V617F mutation \cite{pastore2020prmt5}.

\item[(ii)] PRMT5 inhibition reduces the expression of E2F1. Thus,
  PRMT5 up-regulates E2F1 \cite{pastore2020prmt5}.

\item[(iii)] The expression of E2F1 induces all genes of the endogenous
  MCM family \cite{ohtani1999cell}. E2F1 is a transcriptional
  activator of AURKB \cite{yu2017six3} that can up-regulate AURKB and
  MCM5 expression \cite{reyes2018rna}. Overexpressing E2F1 alone
  results in the up-regulation of MCM5 and TK1
  \cite{koushyar2017prohibitin}. In sum, E2F1 up-regulates AURKB,
  MCM2, MCM4, MCM5, and TK1.

\item[(iv)] From observations (a)-(c), JAK2 mutation indirectly
  up-regulates AURKB, MCM2, MCM4, MCM5, and TK1 through PRMT5 and
  E2F1. We propose that JAK2 mutation can weakly but directly
  down-regulate these genes. This hypothesis can be verified
experimentally by introducing the JAK2 mutation after the knockdown
  or knockout of PRMT5 or E2F1 and observing a decreased expression of
  AURKB, MCM2, MCM4, MCM5, and TK1.

\item[(v)] We propose that a mutated TET2 can down-regulate E2F1
  directly, or indirectly through PRMT5. This down-regulation cancels
  out the up-regulation JAK2 $\to$ PRMT5 $\to$ E2F1. This means E2F1 (and
  possibly PRMT5) expression satisfies $y^*_\jm > y^*_\jt$ and
  $y^*_\om>y^*_\tm$.

\end{itemize}

In summary, in patients without a TET2 mutation, the JAK2 mutation can
up-regulate PRMT5 and E2F1, which in turn up-regulate AURKB, MCM2,
MCM4, MCM5, and TK1; this strong indirect up-regulation of JAK2 $\to$
PRMT5 $\to$ E2F1 $\to$ AURKB/MCM2/MCM4/MCM5/TK1 can cover the weak
direct down-regulation JAK2 $\dashv$ AURKB/MCM2/MCM4/MCM5/TK1, and the
overall effect is $x^*_\om<x^*_\jm$. In the presence of the TET2
mutation, the up-regulation JAK2 $\to$ PRMT5 $\to$ E2F1 is covered by
the down-regulation TET2 $\dashv$ PRMT5/E2F1; therefore, PRMT5 and
E2F1 are locked to low levels so that the only effective regulation of
JAK2 is the down-regulation JAK2 $\dashv$
AURKB/MCM2/MCM4/MCM5/TK1. This means we have $x^*_\tm > x^*_\tj$.


Ortmann et al. \cite{ortmann2015effect} reported ten genes that follow
$x^*_\om<x^*_\jm$ but also $x^*_\tm > x^*_\tj$: AURKB, FHOD1, HTRA2,
IDH2, MCM2, MCM4, MCM5, TK1, UQCRC1, and WDR34. Our model can explain
five of them (AURKB, MCM2, MCM4, MCM5, TK1) with the same pathway JAK2
$\to$ PRMT5 $\to$ E2F1 $\to$ AURKB/MCM2/MCM4/MCM5/TK1, while the role
of Y can be played by E2F1 and/or PRMT5. Fig.~\ref{f5}(b) shows a
simple gene regulatory network that is consistent with the
observations.  Our proposed model also implies a number of
predictions. Specifically, the JAK2 mutation weakly but directly
down-regulates genes that satisfy $x^*_\om < x^*_\jm$ and $x^*_\tm >
x^*_\tj$. Besides, E2F1 and possibly PRMT5 have $y^*_\jm > y^*_\jt$
and $y^*_\om>y^*_\tm$.

The pathway JAK2 $\to$ PRMT5 $\to$ E2F1 $\to\cdots$ is but one
possibility. There is also evidence for the role of p53 in observation
{\bf (3)}. JAK2 V617F negatively regulates p53
  stabilization \cite{nakatake2012jak2v617f}, while p53 can regulate
  AURKB and MCM5 \cite{reyes2018rna}.  The complete gene regulatory network
should be determined using certain inference methods based on gene
expression data \cite{wang2022inference,bocci2022splicejac}.

\subsection{Model for observation (4)}
\label{p4}
\begin{figure}[h!]
  \centering
    \includegraphics[width=0.38\linewidth]{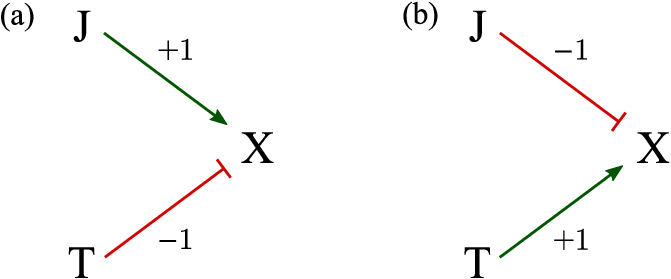}
\caption{\small (a) A schematic of the model $\lambda=2+
  \mathds{1}_{\mathrm{J}}- \mathds{1}_{\mathrm{T}}$ in Eq.~\ref{eqnew}
  which explains $x^*_\tj<x^*_\jt$.  If the effects of JAK2 and TET2
  mutations towards the input $\lambda$ are together greater than
  $0.4$ (\textit{i.e.}, with JAK2 but not TET2), the system is forced
  to be on the high-$x^{*}$ branch; if the contribution to $\lambda$
  input JAK2 and TET2 is smaller than $-0.4$ (\textit{i.e.}, with TET2
  but not JAK2), the system ends up on the low-value branch. (b) The
  model $\lambda=2- \mathds{1}_{\mathrm{J}}+ \mathds{1}_{\mathrm{T}}$
  can yield $x^*_\jt<x^*_\tj$. If the contribution from JAK2 and TET2
  mutations to $\lambda$ is greater than $0.4$ (i.e., with TET2 but
  not JAK2), the system is forced onto the high-$x^{*}$ branch; if the
  JAK2 and TET2 contributions to the input $\lambda$ is smaller than
  $-0.4$ (i.e., with JAK2 but not TET2), the system is forced onto the
  low-$x^{*}$ branch.}
	\label{f7}
\end{figure}

To explain observation {\bf (4)} that TJ and JT have different
effects, namely $x^*_\tj\ne x^*_\jt$, consider Eq.~\ref{eqnew} with
$\lambda=2+ \mathds{1}_{\mathrm{J}}- \mathds{1}_{\mathrm{T}}$.  With J but
not T, $\lambda=3$ and X lies in its only high-value stationary state
$x^*_\jm\approx 3.3$; if T appears after J, then $\lambda=2$, and X
\added{remains in its high-value branch with stationary level
  $x^*_\jt=3$.} If the TET2 mutation arises with a JAK2 mutation,
$\lambda=1$ and the steady-state expression of X is $x^*_\tm\approx 0.7$; if
J appears after T, then $\lambda=2$ and X expression remains in its
low-value branch with stationary value $x^*_\tj=1$. See Fig.~\ref{c0}
for a more detailed description. For MPN patients, if the order is JT,
the final X expression is high ($x^*_\jt=3$); if the order is TJ, the
final X expression level is low ($x^*_\tj=1$). See Fig.~\ref{f7}(a) for
an illustration of this model explaining $x^*_\jt>x^*_\tj$.
%
%

To explain $x^*_\jt<x^*_\tj$, consider Eq.~\ref{eqnew} with $\lambda=2-
\mathds{1}_{\mathrm{J}}+ \mathds{1}_{\mathrm{T}}$. If the mutation
order is JT, the final X expression level is low ($x^*_\jt=1$); if the
order is TJ, the final X expression level is high ($x^*_\tj=3$).
This regulation control mechanism is illustrated in 
Fig.~\ref{f7}(b).


\section{Models for non-commutativity in cell population and age}
\label{at}
In this section, we build models to explain observations {\bf(5, 6)}
that the age of diagnosis and the populations of cancer cells 
depend on the order of the two mutations experienced by the patient. 
Clinically, the mutations are not commutative.

\subsection{Different mechanisms for explaining observations (5, 6)}

For observations {\bf (5, 6)}, the cell population and age are
measured at the time of diagnosis. However, it is difficult to know
the time interval between acquiring the second mutation and diagnosis
or to model the disease progression during this time. Therefore, we
analyze observations {\bf (5, 6)} focussing on the time the first
double-mutation cell (with both JAK2 and TET2 mutations) appears.
Doing so, we must assume an amendment to the observations:

\vspace{3mm}
\noindent \textbf{(5')} For TET2-first patients, at the time when the
first TET2-JAK2 cell appears, the percentage of TET2-only cells is
significantly higher than the percentage of JAK2-only cells at the
time when the first JAK2-TET2 cell appears in JAK2-first patients.
\vspace{2mm}

\noindent \textbf{(6')} For JAK2-first patients, the time at which the
first JAK2-TET2 cell appears is significantly earlier than the time at
which the first TET2-JAK2 cell appears for TET2-first patients.
\vspace{4mm}

Thus, we are assuming that the time delay between the appearance of
the first double-mutation cell and diagnosis is independent of the
order of mutation. In this case, observation \textbf{(6)} and
observation \textbf{(6')} are equivalent.

The relationship between observation \textbf{(5)} and observation
\textbf{(5')} is complicated. After the appearance of one mutation,
gene expression levels can reach the new stationary states relatively
quickly, \added{typically within a cell lifespan.} However,
\textit{populations of cells} with different mutations may take years
before reaching steady-state numbers (\textit{e.g.}, for cells with
both JAK2 and TET2 mutations to dominate). Therefore, between the
appearance of the first double-mutation cell and diagnosis, the cell
population composition may have changed significantly.  Nevertheless,
we assume that the percentage of JAK2-only cells in JAK2-first
patients and the percentage of TET2-only cells in TET2-first patients
does not change appreciably before diagnosis.  In this sense,
observations \textbf{(5)} and \textbf{(5')} can be assumed equivalent.

\deleted{Since MPN is the result of certain mutations, to simplify the
discussion, we assume that diagnosis is perfect and that the time of
diagnosis is coincident with the time the first cell with both JAK2
and TET2 mutations appears.  Observations {\bf(5, 6)} can be
  further refined:}

\deleted{Also, we assume that the percentage of cells with only one
  mutation is similar for the time when the first cell with both JAK2
  and TET2 mutations appears and the time when this percentage is
  measured.  Therefore, we actually discuss two other observations:}


\vspace{3mm}

\noindent \textbf{(A)} Ortmann et al. \cite{ortmann2015effect} propose
that \emph{cells with a JAK2 mutation have only a mild proliferation
  advantage while cells with a TET2 mutation (whether JAK2 is present
  or not) have a more significant proliferation advantage}. This
feature would explain observation ({\bf 5'}). If the JAK2 mutation
first appears, such JAK2-only cells proliferate only slightly faster
than non-mutant cells. Thus, there are few JAK2-only cells that can
acquire the TET2 mutation.  If the TET2 mutation appears first, such
TET2-only cells grow much faster than non-mutant cells. Thus, there is
a higher population of TET2-only cells when the JAK2 mutation
appears. In the following simulations, we find that this mechanism can
also be used to explain observation ({\bf 6'}).  The model by Teimouri
and Kolomeisky \cite{teimouri2021temporal} is relevant to this
mechanism in that they assume different proliferation rates between
JAK2-only mutated cells and TET2-only mutated cells, but assume equal
proliferations rates for JT and TJ cells. They incorporate a number of
assumptions that are not satisfied in this system.


\vspace{3mm}

\noindent \textbf{(B)} Since different mutations generally appear with
different rates \cite{lynch2010evolution}, a more general model can
also include different rates for the different mutations.  Here, we
propose a mechanism in which \emph{JAK2 and TET2 have different
  mutation rates}, explaining observations {\bf (5')} and {\bf (6')}.
If the mutation rate of JAK2 is lower than that of TET2, then when
JAK2 appears first, JAK2-only cells have a shorter time to proliferate
before a TET2 mutation appears. This provides a basis for observation
{\bf (5')}. The explanation for observation {\bf (6')} is given in
Subsection~\ref{s5.4} below.
\vspace{3mm}

\noindent \textbf{(C)} We propose a cooperative mutation mechanism
that can also lead to {\bf (5')} and {\bf (6')}: cells with the JAK2
mutation carry a higher mutation rate for TET2 mutation.  In other
words, \emph{a JAK2 mutation can induce an additional TET2 mutation}.
Therefore, a TET2 mutation can arise quickly after the first JAK2
mutation appears, explaining observation {\bf (6')}. JAK2-only cells do
not have much time to proliferate (into JAK2-only daughter cells)
before the appearance of a TET2 mutation, consistent with observation
{\bf (5')}.

\subsection{Generalized Moran process}

\begin{figure}[h!]
  \centering
    \includegraphics[width=0.5\linewidth]{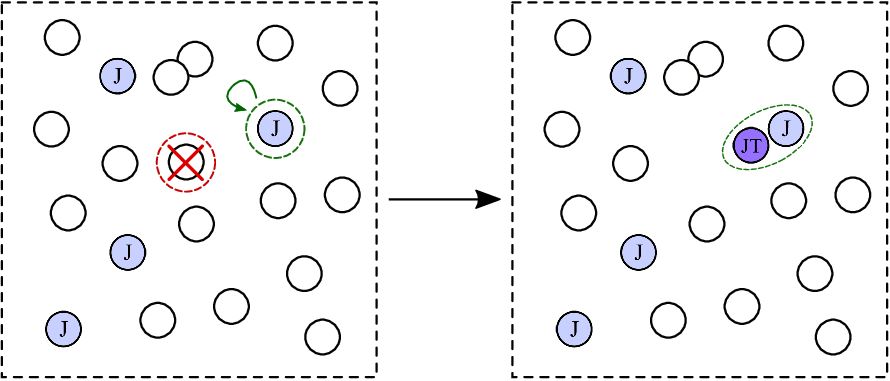}
  \caption{\small A schematic of the steps in our Moran process. At
    some time, the system contains sixteen wild-type cells and four
    JAK2-mutated cells.  In each timestep, one cell (wild-type) is
    chosen for removal (red-dashed circle), while another (J) is
    chosen for replication (green-dashed circle), during which one
    daughter may acquire a mutation.  In this example, a J cell
    divides into a J cell and a double-mutant JT cell, thus defining
    the end point of our simulation.}
	\label{MORAN}
\end{figure}

We implement a simple Moran population model to explore consequences
of mechanisms {\bf (A)}, {\bf (B)}, and {\bf (C)}.  Cell population
dynamics that include state transitions have been widely studied
\cite{zhou2014multi,niu2015phenotypic,chen2016overshoot,angelini2022model}. To
mathematically model observations {\bf (5')} and {\bf (6')}, we
consider a simple discrete-time Moran model
\cite{fudenberg2004stochastic,quan2011evolutionary}, shown in
Fig.~\ref{MORAN}, for cell
populations that include mutations. A continuous-time Moran model can
also be straightforwardly constructed and analyzed. A related
two-mutation branching process has formulated describe the first times
to acquire double mutations, but did not distinguish the order of
mutation acquisition \cite{CHOU_JTB}. Moreover, unlike branching
processes \cite{jiang2017phenotypic}, the total number of cells is
fixed in our Moran process. This is a reasonable approximation for
stable hematopoietic stem cell populations and allow us to easily
estimate relative populations of all cells.  We will assume that cells
can exist in five states: non-mutant, JAK2-only, TET2-only, JAK2-TET2,
and TET2-JAK2. Here, cells with two mutations, for example JAK2-TET2,
are those that are part of a lineage that was started when a single
JAK2-mutation mother cell divided into daughter cells that acquired
the TET2 mutation.

In the following, the suffix $\om$ denotes wild-type cells, $\jm$
denotes JAK2-only cells, $\tm$ describes TET2-only cells, $\jt$
defines JAK2-TET2 cells, and $\tj$ labels TET2-JAK2 cells.  The number
of wild-type cells is $n_\om$ and the relative birth and death
coefficients of these unmutated cells are $b_\om$ and $d_\om$,
respectively.  The analogous populations and birth and death
coefficients are similarly defined for $\jm$, $\tm$, $\jt$, $\tj$-type
cells.
At each time step, one cell is randomly--weighted by the death rate of
its type--picked for removal. Simultaneously, another cell is
randomly--weighted by its birth rate--picked for replication. After
division, one daughter cell will remain in the same state as the
mother cell, while the other may transform into another type according
the corresponding mutation probability.  There are four possible
mutation probabilities $m_{\om \to \jm}$, $m_{\om \to \tm}$, $m_{\jm
  \to \jt}$, and $m_{\tm \to \tj}$.  For example, $m_{\om \to \jm}$ is
the probability that the chosen daughter cell of a wild-type mother
cell acquires the JAK2 mutation. The state space of this process is
$(n_\om, n_\jm, n_\tm, n_\jt, n_\tj)$, with a total fixed-population
constraint $\sum_{j}n_{j} =n$.

At each time point, the probability that a wild-type cell is chosen
for elimination is
\begin{equation*}
	\begin{split}
& \mathbb{P}[(n_\om,n_\jm,n_\tm,n_\jt,n_\tj)\to (n_\om-1,n_\jm,n_\tm,n_\jt,n_\tj)]\\
& \hspace{2cm} =  \frac{n_\om d_\om}{n_\om d_\om+n_\jm d_\jm+n_\tm d_\tm+n_\jt d_\jt+n_\tj d_\tj}.
	\end{split}
\end{equation*}
The probability of selecting other cell types for death are similarly
defined. The probability that a wild-type cell is chosen to divide,
and that no mutations arise in the daughter cells is
\begin{equation*}
	\begin{split}
&\mathbb{P}[(n_\om,n_\jm,n_\tm,n_\jt,n_\tj)\to (n_\om+1,n_\jm,n_\tm,n_\jt,n_\tj)]\\
 & \hspace{2cm} = \frac{n_\om b_\om (1-m_{\om \to \jm}-m_{\om \to \tm})}{n_\om b_\om+n_\jm b_\jm+n_\tm b_\tm+n_\jt b_\jt+n_\tj b_\tj}.
	\end{split}
\end{equation*}
Similarly, the probabilities of generating additional cells of other cell types are
\begin{equation*}
	\begin{split}
		&\mathbb{P}[(n_\om,n_\jm,n_\tm,n_\jt,n_\tj)\to (n_\om,n_\jm+1,n_\tm,n_\jt,n_\tj)]\\
		& \hspace{2cm} = \frac{n_\om b_\om m_{\om \to \jm}}{n_\om b_\om+n_\jm b_\jm+n_\tm b_\tm+n_\jt b_\jt+n_\tj b_\tj}\\
	  &  \hspace{2.6cm} +\frac{n_\jm b_\jm(1-m_{\jm \to \jt})}{n_\om b_\om+n_\jm b_\jm+n_\tm b_\tm+n_\jt b_\jt+n_\tj b_\tj},
	\end{split}
\end{equation*}

\begin{equation*}
	\begin{split}
		&\mathbb{P}[(n_\om,n_\jm,n_\tm,n_\jt,n_\tj)\to (n_\om,n_\jm,n_\tm+1,n_\jt,n_\tj)]\\
		& \hspace{2cm} = \frac{n_\om b_\om m_{\om \to \tm}}{n_\om b_\om+n_\jm b_\jm+n_\tm b_\tm+n_\jt b_\jt+n_\tj b_\tj}\\
		&  \hspace{2.6cm} +\frac{n_\tm b_\tm (1-m_{\tm \to \tj})}{n_\om b_\om+n_\jm b_\jm+n_\tm b_\tm+n_\jt b_\jt+n_\tj b_\tj},
	\end{split}
\end{equation*}

\begin{equation*}
	\begin{split}
		&\mathbb{P}[(n_\om,n_\jm,n_\tm,n_\jt,n_\tj)\to (n_\om,n_\jm,n_\tm,n_\jt+1,n_\tj)]\\
		&  \hspace{2cm}  = \frac{n_\jm b_\jm m_{\jm \to \jt}}{n_\om b_\om+n_\jm b_\jm+n_\tm b_\tm+n_\jt b_\jt+n_\tj b_\tj}\\
		&  \hspace{2.6cm} +\frac{n_\jt b_\jt}{n_\om b_\om+n_\jm b_\jm+n_\tm b_\tm+n_\jt b_\jt+n_\tj b_\tj},
	\end{split}
\end{equation*}
and

\begin{equation*}
	\begin{split}
		&\mathbb{P}[(n_\om,n_\jm,n_\tm,n_\jt,n_\tj)\to (n_\om,n_\jm,n_\tm,n_\jt,n_\tj+1)]\\
	& \hspace{2cm} = \frac{n_\tm b_\tm m_{\tm \to \tj}}{n_\om b_\om+n_\jm b_\jm+n_\tm b_\tm+n_\jt b_\jt+n_\tj b_\tj}\\
		&  \hspace{2.6cm} +\frac{n_\tj b_\tj}{n_\om b_\om+n_\jm b_\jm+n_\tm b_\tm+n_\jt b_\jt+n_\tj b_\tj}.
	\end{split}
\end{equation*}	


\subsection{Simulation results}

\begin{table}[]
  \begin{center}
	\begin{tabular}{l|lllll|llll}
	  & $b_\om$  & $b_\jm$  & $b_\tm$  &$b_\jt$   & $b_\tj$   &     $m_{\om \to \jm}$   & $m_{\om \to \tm}$    &  $m_{\jm \to \jt}$   & $m_{\tm \to \tj}$    \\
          \hline
		Mechanism {\bf (A)}   & 1 & 2 & 4 & 4 & 4      & 0.1 & 0.1 & 0.1 & 0.1 \\
		Mechanism {\bf (B)}  & 1 & 2 & 2 & 2 & 2     & 0.1 & 0.2 & 0.2 & 0.1 \\
		Mechanism {\bf (C)} & 1 & 2 & 2 & 2 & 2     & 0.1 & 0.1 & 0.2 & 0.1
	\end{tabular}
        \end{center}
	\caption{\added{A generalized Moran process is simulated to
            explore three different mechanisms, {\bf (A)}, {\bf (B)},
            and {\bf (C)}, with corresponding relative birth
            coefficients $b$ and mutation probabilities $m$
            listed. The mutation probability is the probability that
            one daughter acquires a mutation at birth.}}
	\label{para}
\end{table}

In the simulations, for all three mechanisms, we set the initial
population $n_\om=100,n_\jm=n_\tm=n_\jt=n_\tj=0$, so that the total
population is $n=100$. We set a common value for all death
probabilities $d_\om=d_\jm=d_\tm=d_\jt=d_\tj=1$. Table~\ref{para}
lists the relative birth rates and mutation probabilities associated
with three different mechanisms. 

In Mechanism {\bf (A)}, all mutations have the same probability and
cells with the TET2 mutation proliferate faster than those with
JAK2-only mutations. In Mechanism {\bf (B)}, all cells with at least
one mutation have the same relative birth coefficients, while the
appearance probability of a TET2 mutation is set higher than that of a
JAK2 mutation. In Mechanism {\bf (C)}, all cells with at least one
mutation have the same relative birth coefficients, but a cell with a
JAK2 mutation more likely acquires a TET2 mutation upon division,
corresponding to a larger $m_{\jm \to \jt}$.

Since we use this Moran process to study observations ({\bf 5', 6'})
at the time that the first double-mutation cell appears, the process
is stopped once $n_\jt=1$ or $n_\tj=1$. At this point, if both
$n_\jm>0$ and $n_\tm>0$ the order of mutation cannot be inferred and
this simulation result is abandoned. If $n_\jt=1$ and $n_\tm=0$, we
record the corresponding $n_\jm$ and the current time point $T$. This
mechanism reflects a JAK2-first patient. If $n_\tj=1$ and $n_\jm=0$,
we record the corresponding $n_\tm$ and the current time point
$T$. This mechanism reflects a TET2-first patient.  For each
mechanism, $10^{6}$ trajectories were simulated.  Although most
trajectories are abandoned, at least $10^{4}$ trajectories remained
from which JAK2-first and TET2-first dynamics could be determined and
sufficient statistics extracted.

To verify observation ({\bf 5'}), we compare $\mathbb{E}(n_\jm\mid
n_\jt=1,n_\tm=0)$ and $\mathbb{E}(n_\tm\mid n_\tj=1,n_\jm=0)$. To
investigate observation ({\bf 6'}), we compare $\mathbb{E}(T\mid
n_\jt=1,n_\tm=0)$ to $\mathbb{E}(T\mid n_\tj=1,n_\jm=0)$.  We also use
a $t$-test to examine whether the difference in mean cell populations
are significant. For each scenario, we run the simulation $10^6$
times.

For Mechanism {\bf (A)}, $\mathbb{E}(n_\jm\mid n_\jt=1,n_\tm=0)= 4.02
< 5.41=\mathbb{E}(n_\tm\mid n_\tj=1,n_\jm=0)$, and the $p$-value from
the $t$-test is smaller than $10^{-200}$. We also find
$\mathbb{E}(T\mid n_\jt=1,n_\tm=0)= 24.15 < 25.93=\mathbb{E}(T\mid
n_\tj=1,n_\jm=0)$, with a $t$-test $p$-value $\sim 10^{-40}$.  This
scenario can generate observation ({\bf 6'}). Since $b_\tm>b_\jm$, the
probability that $n_\jt=1,n_\tm=0$ is smaller than that of
$n_\tj=1,n_\jm=0$. Simulations with $n_\jt=1,n_\tm=0$ generally mean
that the JAK2 mutation happens to arise more quickly, and that these
JAK2-only cells happen to divide more frequently.
Under Mechanism {\bf (B)}, $\mathbb{E}(n_\jm\mid n_\jt=1,n_\tm=0)=
2.37 < 6.51=\mathbb{E}(n_\tm\mid n_\tj=1,n_\jm=0)$ and
$\mathbb{E}(T\mid n_\jt=1,n_\tm=0)= 11.52< 24.75=\mathbb{E}(T\mid
n_\tj=1,n_\jm=0)$, both with a $t$-test $p$-value less $10^{-200}$.
Finally, in Mechanism {\bf (C)}, $\mathbb{E}(n_\jm\mid
n_\jt=1,n_\tm=0)= 3.59 < 4.27=\mathbb{E}(n_\tm\mid n_\tj=1,n_\jm=0)$
with $p$-value $\sim 10^{-192}$ and $\mathbb{E}(T\mid
n_\jt=1,n_\tm=0)= 22.79 < 25.97=\mathbb{E}(T\mid n_\tj=1,n_\jm=0)$
with $p$-value $\sim 10^{-133}$.
After applying the Bonferroni correction to these six tests, we find
the probability of rejecting at least one of the above results is
$\alpha=10^{-39}$.

In this model, we see that all three mechanisms can produce
observations ({\bf 5', 6'}).  Biologically, it is natural to assume
that JAK2 and TET2 have different mutation rates (Mechanism {\bf
  (B)}). Mechanisms {\bf (A)} and {\bf (C)} require more supporting
evidence, so we primarily propose the mechanism associated with {\bf
  (B)}, which is sufficient to explain observations.

\subsection{Theoretical analysis of Mechanism {\bf (B)} for observation {\bf (6')}}
\label{s5.4}

A generalized Moran process model is difficult to study
analytically. To explain why Mechanism {\bf (B)} produces observation
({\bf 6'}), we consider a simplified model, which is a limiting
mechanism of the generalized Moran model.

Assume $b_\jm=b_\tm=b_\jt=b_\tj\gg b_\om$. This means that once one
mutation appears, cells with this mutation will dominate the
population. Thus, the situation in which both populations are
appreciable, $n_\jm>0$ and $n_\tm>0$, does not arise. If we further
assume $m_{\om \to \jm}=m_{\tm \to \tj}\equiv m_1$, $m_{\om \to
  \tm}=m_{\jm \to \jt}\equiv m_2$ and that $m_1< m_2 \ll 1$, we can
approximate the distribution of times (number of time steps in our
discrete-time simulations) $T_1$ for JAK2 mutation to appear by an
exponential distribution with parameter $m_1$ so that $\mathbb{E}(T_1)
\approx 1/m_1$.  The time $T_{2}$ for a TET2 mutation to appear is
also exponentially distributed with parameter $m_2$ so that
$\mathbb{E}(T_2) \approx 1/m_2$, and $T_1$ and $T_2$ are approximately
independent. $T_1<T_2$ corresponds to the JAK2-first scenario, while
$T_1>T_2$ results in a TET2-first observation. Now, assume a faster
TET2 mutation rate $(m_2 > m_1)$ and define $T=\max\{T_1,T_2\}$ as the
time at which both mutations first arise. We find, approximating
in continuous-time, 

\begin{equation}
\begin{aligned}
\mathbb{E}(T\mid T_1>T_2) \approx &\, \frac{\displaystyle\int_{0}^{\infty} 
 \int_{0}^{t_1}t_1 m_1e^{-m_1 t_1}
  m_2 e^{-m_2 t_2}\mathrm{d}t_2\mathrm{d}t_1}{\displaystyle \int_{0}^{\infty} 
\int_{0}^{t_1}m_1 e^{-m_1 t_1} m_2 e^{-m_2 t_2}
\mathrm{d}t_2\mathrm{d}t_1}=
\frac{2m_1 m_2+m_2^2}{m_1 m_2(m_1+m_2)}, \\[4pt]
\mathbb{E}(T\mid T_1< T_2) \approx &\, \frac{\displaystyle \int_{0}^{\infty} \int_{0}^{t_2}t_2
m_1 e^{-m_1 t_1}m_2 e^{-m_2 t_2}\mathrm{d}t_1\mathrm{d}t_2}
{\displaystyle \int_{0}^{\infty} \int_{0}^{t_2}m_1 e^{-m_1 t_1}_2 e^{-m_2 t_2}\mathrm{d}t_1\mathrm{d}t_2}
=\frac{m_1^2+2m_1 m_2}{m_1 m_2 (m_1+m_2)}.
\end{aligned}
\end{equation}
Since $m_2>m_1$, $\mathbb{E}(T\mid T_1>T_2)>\mathbb{E}(T\mid T_1 <
T_2)$. For JAK2-first patients ($T_1< T_2$), the waiting time for both
mutations to appear is shorter than that for TET2-first patients
($T_1>T_2$). \added{One explanation for this is that when $T_1<T_2$,
  it is more likely that $T_1$ is exceptionally shorter, not that
  $T_2$ is exceptionally longer.}


\section{Discussion and Conclusions}
\label{disc}
In this paper, we consider two genetic mutations in MPN: JAK2 and
TET2. The effect of one mutation depends on whether the other mutation
is present. When both the mutations are present, the order of their
appearance also affects gene expression. For MPN, the order of the
JAK2 V617F and DNMT3A mutations can also affect cellular proliferation
\cite{nangalia2015dnmt3a}. \added{The TET2 and DNMT3A mutations confer
  epigenetic changes in transcription that are passed on to daughter
  cells, thus providing a mechanism of ``memory'' required for
  bi/multistability and ultimately an order-of-mutation effect.}
Dependence of cell populations on the order of mutation also appear in
other types of cancer. For example, in adrenocortical carcinomas, if
the Ras mutation appears before the p53 mutation, the tumor will be
malignant and metastatic, but if the p53 mutation appears before the
Ras mutation, the tumor will be benign
\cite{herbet2012acquisition}. Similar observations can be found in
other contexts
\cite{levine2019roles,turajlic2018deterministic,caravagna2018detecting}.

We constructed several sub-models to explain the features of
order-of-mutation effects in cancer, specifically addressing
observations recorded to date for the JAK2/TET2 mutation pair in
MPN. In Subsection~\ref{p3}, we describe experimental evidence that
partially verifies our model. We also provided conjectures that can be
tested experimentally: JAK2 mutation can weakly down-regulate AURKB,
MCM2, MCM4, MCM5, and TK1 directly; TET2 mutation can down-regulate
E2F1 and/or PRMT5.

\added{Although we have developed a mathematical framework consistent
  with all observations to date, there are other possible processes
  that can lead to the rich set of observations discussed.
  Potential interactions with the adaptive immune system may inhibit
  cancer progression
  \cite{mellman2011cancer,altrock2015mathematics}. Cancer may also
  inhibit the proliferation of white blood cells
  \cite{hamanishi2007programmed}, which can lead to multistability in
  mathematical models of immune response to cancer
  \cite{garcia2020cancer,li2017bistability,vithanage2021bistability}.
  Since certain mutations can help cancer cells escape the immune
  system \cite{hanahan2011hallmarks}, it is possible that the order of
  mutations affects cancer cell populations indirectly by interfering
  with the immune system. Finally, cancer cells can also affect and be
  affected by their microenvironments and other cells (through
  \textit{e.g.}, epigenetically driven ``microenvironment
  feedback''). These nonlinear interactions have been modeled can lead
  to nonlinear dynamics in relative populations of different cancer
  cell types (different epigenetic or mutational states)
  \cite{GOYAL_PRE}.  Further developing models that incorporate immune
  and indirect cell-cell interactions could potentially lead to
  non-additivity and non-commutivity of mutation order in both gene
  expression and cell population dynamics.  Formulating such
  mathematical frameworks, especially those coupling intracellular
  state dynamics to proliferating cell population will be the subject
  of future investigation.}



\subsection*{Acknowledgements} YW and TC acknowledge support
from the National Institutes of Health through grant R01HL146552.

\appendix
\section{Detailed review of previously known models}
\label{review}

Ortmann et al. \cite{ortmann2015effect} assume that TET2 mutation can
significantly increase the proliferation rate of cancer stem cells,
while JAK2 mutation only has a weak growth advantage. Therefore, for
TET2-first patients, TET2-only cells first spread, and TET2-JAK2 cells
(which do not have a significant growth advantage over TET2-only
cells) do not dominate. For JAK2-first patients, JAK2-only cells do
not spread that much, while JAK2-TET2 cells (after they appear) can
dominate. Therefore, TET2-first patients have a much higher percentage
of cells with only one mutation, consistent with observation {\bf
  (5)}. In Section~\ref{at}, we also discuss that these assumptions
can explain observation {\bf (6)}. See also the interpretation by
Swanton \cite{swanton2015cancer}.

Kent and Green \cite{kent2017order} propose two explanations for
observation {\bf (4)}.\\

{\noindent (i) Both JAK2 and TET2 mutations can participate in
	epigenetic regulation \cite{dawson2009jak2,ito2010role,shih2012role},
	but the regulation mechanism might be incompatible. For example, the
	first mutation might lead to the occlusion of certain genomic regions,
	so that the second mutation cannot regulate genes in those
	regions. This mechanism would lead to $x^*_\jt = x^*_\jm \ne x^*_\tm =
	x^*_\tj$, which is not consistent with other observations.}
%
\\

\noindent (ii) For either JAK2-first or TET2-first patients, before
the appearance of the second mutation, different first mutations might
lead to different cell types and abundances, leading to different 
microenvironments in which double mutant cells 
that subsequently arise find themselves. This indirect effect 
can also shape disease progression.


In a related study, Roquet et al. \cite{roquet2016synthetic}
consider the effect of recombinases (i.e., genetic recombination
enzymes) on gene sequences. When applying different recombinases to
gene sequences, their order of application can lead to different
results. For example, consider a gene sequence $12312$ and two
recombinases \texttt{A}, and \texttt{B}. Suppose \texttt{A} deletes
genes between the ``$1$s'', and \texttt{B} inverts genes
between the two ``$2$s'' (if there are not two ``$2$s'', \texttt{B} does
nothing). If the DNA is exposed to \texttt{A} before \texttt{B},
the gene sequence becomes
\[12312\xrightarrow[]{\texttt{A}}112\xrightarrow[]{\texttt{B}}112.\] 
If \texttt{A} is added after \texttt{B}, the gene sequence becomes
\[12312\xrightarrow[]{\texttt{B}}12132\xrightarrow[]{\texttt{A}}1132.\] 
Although this system describes rearrangements and not specific
mutations, it nonetheless provides a possible mechanism for
observation {\bf (4)}.

Ascolani and Li{\`o} \cite{ascolani2019modeling} constructed a complex
cellular automata model to describe cancer metastasis. They assume
that different mutations lead to different proliferation and/or
apoptosis rates. This assumption, as was invoked by Ortmann et al.,
can explain observation {\bf (5)}. However, although Ascolani and
Li{\`o} also take into account how different \textit{orders} of
different mutations can affect proliferation and apoptosis
differently, they did not explicitly apply their model to the observations 
associated with different orders of JAK2 and
TET2 mutations in the MPN system.

Clarke et al. \cite{clarke2019using} model the gene regulatory network
as a generalized boolean network that evolves under certain rules and
exhibits fixed points and/or limit cycles {for gene expression
  levels}. Each mutation fully activates or inhibits one gene (the
  expression level is fixed, similar to the do-operator used in causal
  inference \cite{benferhat2007possibilistic}), thus changing
  the fixed points and/or limit cycles. Certain
combinations of mutations lead to higher
proliferation rates or apoptosis rates, making these mutation patterns
more likely or less likely, respectively. There might be multiple
fixed points and/or limit cycles, and different orders of mutations
might lead to different final states of gene
  expression. Different sequences of perturbations leading to
  different states have been hypothesized for different physiological
  dynamics, including neuroendocrine stress response \cite{LUKIM,CHENG}.
This type of model can be used to explain observations {\bf (4)} and
{\bf (5)}.

Talarmain et al. \cite{talarmain2022hoxa9,talarmain2021modelling}
apply the model in Clarke et al.'s \cite{clarke2019using} paper to the
JAK2/TET2 mutation order problem to explain observation {\bf
  (4)}. They find a concrete generalized boolean network of gene
expression. Mutations can affect the dynamics of this network. When
there is no mutation or just one mutation, the system has one stable
fixed point. When both JAK2 and TET2 mutations are present, the system
has two stable fixed points. Different orders of mutations lead to
different fixed points. The Talarmain et al. model invokes a third,
downstream gene HOXA9 which is directly affected by different orders
of JAK2 and TET2 mutations, which then affects many other downstream
genes.

Mazaya et al. \cite{mazaya2020effects} use a boolean network to model
the effect of mutations. The model dynamics and the explanation of
observation {\bf (4)} are similar to that of Clarke et al.'s model. Mazaya et
al. further analyze this model to study when the network is more
sensitive to different orders of mutations.

Finally, Teimouri and Kolomeisky \cite{teimouri2021temporal} use a
random walk model to study the acquisition of two mutations. The first
stage is a random walk on $0,1,\ldots,n$, representing the number of
cells with the first mutation. The first stage starts at $0$, and
finishes when reaching $n$, meaning that all $n$ cells have the first
mutation. The process terminates if reaching $0$ again before reaching
$n$. The second stage is a random walk on $n,n+1,\ldots,2n$,
representing the number of cells with the second mutation plus
$n$. The second stage starts at $n$, and finishes when reaching $2n$,
meaning that all cells have both mutations. If the process reaches
$2n$, we count the total time and take the expectation. They prove
that if the first mutation has a higher fitness than the second
mutation, then the tumor formation probability is higher, but the time
for tumor formation is longer. This explains observation {\bf (6)}.

\section{Markov chain model for observation {\bf (4)}}
\label{mcm}
In the nonlinear ODE model for observation {\bf (4)}, a cell with no
mutation and a cell with both mutations carry the same gene expression
landscape. \added{Since gene expression at the single-cell level is
  essentially stochastic, we can also build an \textit{ad hoc} Markov
  chain model in which the landscape of gene expression changes with
  the appearance of each mutation.}

In a single cell, the expression level (protein or mRNA count) of gene
X is a random variable $X$ defined on $0,1,2,\ldots$. Define
\[A_1(x,c_1,\mu_1,\sigma_1)=\frac{c_1}{\sqrt{2\pi\sigma_1^2}}e^{\frac{(x-\mu_1+1)^2}{2\sigma_1^2}},\]
\[A_2(x,c_1,\mu_1,\sigma_1)=\frac{c_1}{\sqrt{2\pi\sigma_1^2}}e^{\frac{(x-\mu_1)^2}{2\sigma_1^2}},\]
\[B_1(x,c_2,\mu_2,\sigma_2)=\frac{c_2}{\sqrt{2\pi\sigma_2^2}}e^{\frac{(x-\mu_2+1)^2}{2\sigma_2^2}},\]
\[B_2(x,c_2,\mu_2,\sigma_2)=\frac{c_2}{\sqrt{2\pi\sigma_2^2}}e^{\frac{(x-\mu_2)^2}{2\sigma_2^2}}.\]
$X$ follows a continuous-time Markov chain on $0,1,2,\ldots$ where the
transition rate from $x$ to $x+1$ is
\[r_{x\to x+1}=A_1(x)+B_1(x),\]
and the transition rate from $x+1$ to $x$ is
\[r_{x+1\to x}=A_2(x)+B_2(x).\]
In all cases, we set $\mu_1=1000$, $\mu_2=2000$. If the JAK2 mutation
is not present, we set $c_1=1$ and $\sigma_1=400$ in
$A_1(x,c_1,\mu_1,\sigma_1)$ and $A_2(x,c_1,\mu_1,\sigma_1)$;
otherwise, when the JAK2 mutation is present, we set $c_1=5$ and
$\sigma_1=80$. Similarly, if TET2 mutation is not present, we set
$c_2=1$ and $\sigma_2=400$ in $B_1(x,c_2,\mu_2,\sigma_2)$ and
$B_2(x,c_2,\mu_2,\sigma_2)$; otherwise, when the TET2 mutation is
present, we set $c_2=5$ and $\sigma_2=80$.
\begin{figure}[htb]
	\centering
	\includegraphics[width=0.85\linewidth]{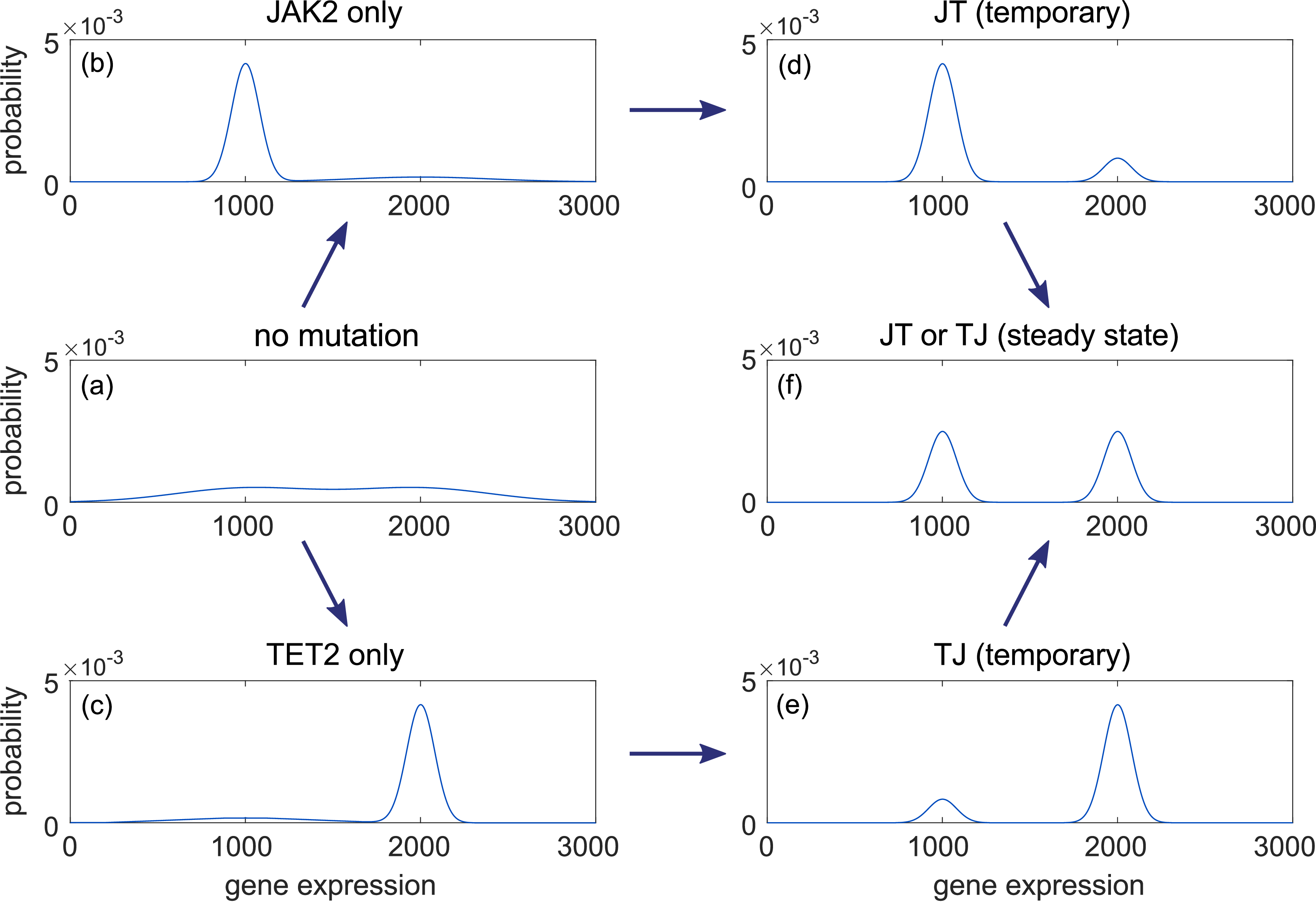}  
	\caption{The probability distribution over gene expression
          levels $x$ under the different enumerated scenarios
          described by the Markov chain model in
          Appendix~\ref{mcm}. (a) The stationary distribution when no
          mutation is present. (b) The distribution of $x$ in cells
          with only a JAK2 mutation. (c) The stationary distribution
          when only TET2 mutation is present. (d) The temporary (but
          long-lived) distribution when a TET2 mutation appears on a
          lineage with an existing JAK2 mutation. (e) Long-lived
          probability distribution of $X$ if the JAK2 mutation
          appears on a background of existing TET2 mutations. (f) The
          final distribution when both JAK2 and TET2 mutations are
          present (regardless of order).}
	\label{mc1}
\end{figure}
\begin{figure}[htb]
	\centering
	\includegraphics[width=0.85\linewidth]{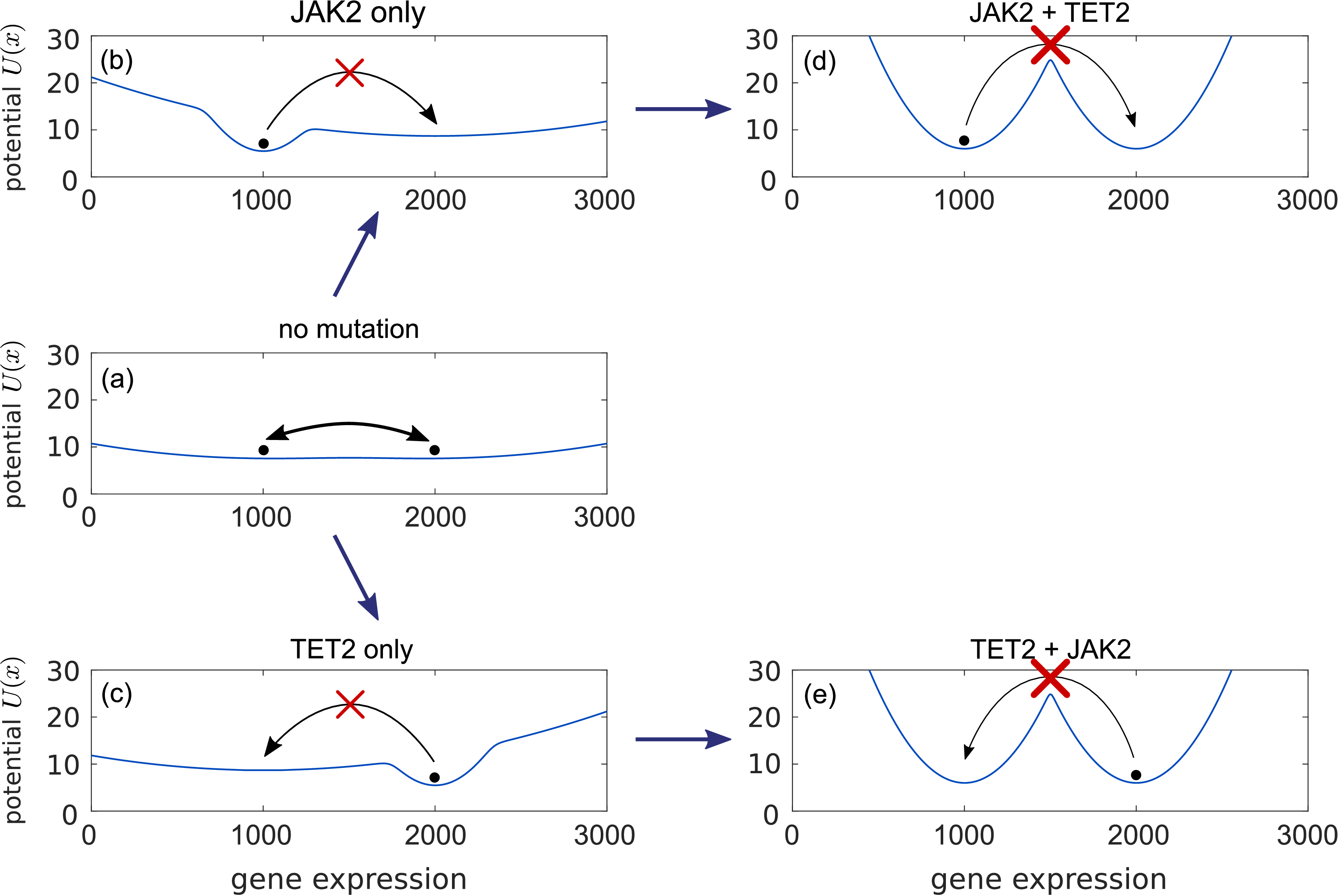}  
	\caption{The effective potential function $U(x)$ of gene
          expression levels $x$ for different scenarios within the
          Markov chain model. (a) The potential function in the
          absence of mutations.  The system can switch between the
          shallow wells near $x=1000$ and $x=2000$. (b) $U(x)$ when
          only the JAK2 mutation is present. The system is confined to
          the deep well near $x=1000$. (c) $U(x)$ when only the TET2
          mutation is present. The system is confined to the deep well
          near the higher expression level $x=2000$. (d) The potential
          function of X gene expression if the TET2 mutation appeared
          after the JAK2 mutation. Since the system was previously
          confined in the well near $x=1000$, it will retain this
          lower level of expression for a long time. (e) $U(x)$ for a
          cell which acquired the JAK2 mutation from a mother cell
          which already had the TET2 mutation. Since the system was
          previously confined in the well near $x=2000$, its gene
          expression will stay high for a long time. Theoretically, in
          the (unrealistically) long time limit, the probability
          densities in both cases (d) and (e) will equipartition
          itself symmetrically across the two equally deep wells.}
	\label{mc2}
\end{figure}

Since this Markov chain has no cycles, the detailed balance condition
is satisfied \cite{wang2020mathematical}, and we can directly
compute the stationary probability distribution $\mathbb{P}(X=x)$
from

\[\mathbb{P}(X=x)r_{x\to x+1}=\mathbb{P}(X=x+1)r_{x+1\to x},\]
which is plotted in Fig.~\ref{mc1}. If no mutation is present, the
stationary distribution is rather flat with two low peaks near
$x=1000$ and $x=2000$, as shown in (a). If only the JAK2 mutation is
present, the stationary distribution is mostly concentrated in a sharp
peak near $x=1000$, with a small flat probability peak near $x=2000$,
as depicted in (b).  If the TET2 mutation appears after the JAK2
mutation, the small flat probability peak near $x=2000$ first sharpens
to a more localized peak near $x=2000$ (d); after an unrealistically
long time (\textit{e.g.}, thousands of years), the heights of two
peaks near $x=1000$ and $x=2000$ equalize, as shown in (f). If only
the TET2 mutation is present, the stationary distribution shown in (c)
is mostly concentrated in a sharp peak near $x=2000$, with a small
nearly flat probability mound near $x=1000$. If the JAK2 mutation then
appears, the small broad probability peak near $x=1000$ first shrinks
to a sharper peak near $x=1000$ (e); after an extremely long time, the
heights of two peaks near $x=1000$ and $x=2000$ again equilibrate as
shown in (f).

We can also define a potential at $X=x$ as the negative logarithm of
the stationary distribution: $U(x)=-\log \mathbb{P}(X=x)$.
Fig.~\ref{mc2} shows the potential function $U(x)$ corresponding to
different temporal configurations of mutations. If no mutation is
present, the potential has two shallow wells near $x=1000$ and
$x=2000$. The expression level can easily move between these two
wells, as shown in Fig.~\ref{mc2}(a). Fig.~\ref{mc2}(b) depicts the
case in which only the JAK2 mutation is present for which there is a
deep well near $x=1000$ and a shallow well near $x=2000$. Here, it is
easy to jump from the shallow well into the deep well, but not the
other way around. Thus, this system is most likely to have expression
level $x=1000$. If the TET2 mutation appeared after the JAK2 mutation
(Fig.~\ref{mc2}(d)), the system will first stay in the deep well near
$x=1000$. Since both wells are deep, there is very little probability
flux from one well to another and the probability distribution relaxes
very slowly (over times longer than the life span of a human) towards
final equipartition.
%
%
If only the TET2 mutation is present, the system is likely to stay in
the deep well near $x=2000$, indicated in Fig.~\ref{mc2}(c). Finally,
if the JAK2 mutations appear after TET2 mutations, the system will
reside in the well near $x=2000$ (see Fig.~\ref{mc2}(e)) before very
slowly becoming equally distributed between $x\approx 1000$ and
$x\approx 2000$.

In this model, different orders of mutations (JT and TJ) lead to the
same final stationary distribution (Fig.~\ref{mc1}(f)). However,
different histories lead to concentration of probabilities to
different wells (Figs.~\ref{mc1}(d) and (e)) on finite timescales. If
this mesoscopic time scale is comparable to the life span of human
being, then the final stationary distribution is \textit{de facto}
inaccessible. This simple probabilistic model can also explain the
difference in gene expression levels between patients with different
orders of mutations.

\bibliographystyle{acm}
\bibliography{Mutation}

\end{document}